\documentclass[final,letter,5p,times,twocolumn]{elsarticle}

\usepackage{amsmath}
\usepackage{amssymb}
\usepackage{graphicx}
\usepackage{mathrsfs}
\usepackage{color}
\usepackage[hypertex]{hyperref}
\hypersetup{colorlinks,citecolor=blue,linkcolor=blue}

\usepackage[T1]{fontenc}
\usepackage[utf8]{inputenc}

\biboptions{sort&compress}
\journal{Elsevier}

\begin{document}

\begin{frontmatter}

\title{Effects of volume corrections and resonance decays on cumulants of net-charge distributions in a Monte Carlo hadron resonance gas model}

\author{Hao-jie Xu}

\ead{haojiexu@pku.edu.cn}

\address{Department of Physics and State Key Laboratory of Nuclear Physics and Technology, Peking University, Beijing 100871, China}

\date{\today}

\begin{abstract} 
	The effects of volume corrections and resonance decays (the
	resulting correlations between positive charges and negative charges) on 
	cumulants of net-proton distributions and net-charge distributions are investigated by using
	a Monte Carlo hadron resonance gas ({\tt MCHRG}) model. The required volume
	distributions are generated by a Monte Carlo Glauber ({\tt MC-Glb}) model. 
	Except the variances of net-charge distributions, the {\tt MCHRG} model with
	more realistic simulations of volume corrections, resonance decays and acceptance cuts can 
	reasonably explain the data of cumulants of net-proton distributions and 
	net-charge distributions reported by the STAR collaboration. The {\tt MCHRG}
	calculations indicate that both the volume corrections and resonance decays
	make the cumulant products of net-charge distributions deviate from the 
	Skellam expectations: the deviations of $S\sigma$ and $\kappa\sigma^{2}$
	are dominated by the former effect while the deviations of $\omega$ are dominated by the
	latter one. 
\end{abstract}

\end{frontmatter}

\section{Introduction}
Recently, tremendous experimental and theoretical efforts have been made
to determine the phase diagram of Quantum Chromodynamics (QCD)~\cite{Jeon:1999gr,Aoki:2006we,Bazavov:2012vg,Gupta:2011wh,Lacey:2014wqa}.
Among various observables, the cumulants of net-charge distributions and net-proton distributions, measured by beam energy scan (BES) program
from the Relativistic Heavy Ion Collider (RHIC) at Brookhaven National
Laboratory (BNL) with a wide range of collisional energies from
$\sqrt{s_{NN}}=7.7$ GeV to $\sqrt{s_{NN}}=200$ GeV~\cite{Adamczyk:2014ipa,Adare:2015aqk,Adamczyk:2014fia,Aggarwal:2010wy,Adamczyk:2013dal}, 
that are expected to provide us some crucial
information about the critical end point (CEP) of QCD phase diagram.
Besides the fluctuation data and theoretical studies on critical fluctuations~\cite{Stephanov:2008qz,Chen:2014ufa,Chen:2015dra,Jiang:2015hri},
the theoretical baselines from non-critical statistical fluctuation studies are also
important~\cite{Cleymans:2004iu,Begun:2004gs,Begun:2006uu,Bzdak:2012an,Xu:2016jaz,Xu:2016qzd,Chatterjee:2009km,Garg:2013ata,Karsch:2010ck,Alba:2014eba,Fu:2013gga,Bhattacharyya:2015zka,BraunMunzinger:2011dn,Karsch:2015zna,Nahrgang:2014fza,Mishra:2016qyj}.

The hadron resonance gas (HRG) models have been employed to investigate the theoretical baselines of multiplicity distributions in various previous studies~\cite{Chatterjee:2009km,Garg:2013ata,Karsch:2010ck,Alba:2014eba,Fu:2013gga,Bhattacharyya:2015zka,BraunMunzinger:2011dn,Karsch:2015zna,Nahrgang:2014fza,Mishra:2016qyj}.
Especially in Ref.~\cite{Nahrgang:2014fza}, Nahrgang and her collaborators split the resonance decay contributions into two parts: an average
part and a probabilistic part, but ignore the correlations generated from resonance decays. This method is
appropriate for the net-proton distributions discussed in Ref.~\cite{Nahrgang:2014fza}, because no correlation between baryons and
anti-baryons can be generated from resonance decays~\cite{Nahrgang:2014fza}. However, the correlations between positive charges and negative charges
should be taken into account in study of the net-charge fluctuations, because lots of positive-negative charge pairs are generated from resonance decays.

For the data reported by the STAR collaboration~\cite{Adamczyk:2014fia}, on the other hand, the volume corrections play significant role on the cumulants (cumulant products)
of net-charge distributions~\cite{Xu:2016jaz,Xu:2016qzd}. Therefore, besides the information of chemical potential and temperature investigated in previous studies, the volume information
is also important for theroetical baseline studies~\cite{Xu:2016qzd}. Meanwhile, with the volume information, one can study the centrality dependence of multiplicity distributions in relativistic heavy ion collisions.

In this work, with the volume information generated by a Monte Carlo Glauber ({\tt MC-Glb}) model, I propose a Monte Carlo hadron resonance gas ({\tt MCHRG}) model to study
the effect of volume corrections and resonance decays on the cumulants of net-proton distributions and net-charge distributions in a transparent way.
Instead of primordial particles used in previous studies, the acceptance cuts can be applied to the decay products in the {\tt MCHRG} simulations.
Based on these advantages, the {\tt MCHRG} model can be used to give more realistic baselines for the cumulants of multiplicity distributions than previous
HRG models.

\section{Monte Carlo hadron resonance gas model} 
\label{main}

In the {\tt{MCHRG}} model, the multiplicities of different particle
species in each event are randomly generated by Poisson distributions,
and the Poisson parameters are calculated by~\cite{Andronic:2005yp}
\begin{equation}
	\lambda_i = V\frac{T_{ch}g_{i}m_{i}^{2}}{(2\pi)^2}\exp{\left(\frac{\mu_{i}}{T_{ch}}\right)}K_{2}\left( \frac{m_{i}}{T_{ch}}\right),
\end{equation}
where $g_{i}$ is degeneracy factor, $m_{i}$ is
particle mass, $T_{ch}$ is chemical freeze-out temperature. The chemical potential
$\mu_{i}=B_{i}\mu_{B}+S_{i}\mu_{S}+Q_{i}\mu_{Q}$  with the baryon
number $B_{i}$, strangeness number $S_{i}$, charge number $Q_{i}$ and the
corresponding chemical potentials $\mu_{B}$, $\mu_{S}$, $\mu_{Q}$. $K$ is modified Bessel function.
Note that I have neglected the effects of quantum statistics on multiplicity fluctuations. 

To study the effect of resonance decays, I use
319 primordial particle species as inputs and 26 stable particle species after
performing Monte Carlo resonance decays, as the particle species
used in~\cite{Nahrgang:2014fza}.
The resonance decay channels are taken from~\cite{Eidelman:2004wy} and the contributions from
weak decays are not taken into account in the present study.

For more realistic simulations of acceptance cuts, the transverse momentum ($p_{T}$) spectra of primordial particles are simulated by the blast-wave model~\cite{Schnedermann:1993ws}
\begin{equation}
	\frac{dN}{p_{T}dp_{T}}  \propto \int_{0}^{1}xdxm_{\perp}I_{0}\left(\frac{p_{\perp}\sinh\rho }{T_{kin}}\right)K_{1}\left(\frac{m_{\perp}\cosh\rho}{T_{kin}}\right),
\end{equation}
where $m_{T}=\sqrt{p_{T}^{2} + m^{2}}$, $T_{kin}$ is kinetic freeze-out temperature, $\rho=\tanh^{-1}(\beta)$ and $\beta=\beta_{S}x$ with $\beta_{S}$ being the velocity at volume boundary. 
And the rapidity ($y$) distribution is modeled by~\cite{Wong:2008ex}
\begin{equation}
	\frac{dN}{dy} \propto \exp{(\sqrt{(L^{2}-y^{2})})}
\end{equation}
with $L = \ln(\sqrt{s_{NN}}/m_{p})$ and $m_p$ is the mass of proton.

To study the centrality dependence of multiplicity distributions, as well as the volume corrections 
on high order cumulants of multiplicity distribution in {\tt{MCHRG}} model, a Monte Carlo Glauber ({\tt{MC-Glb}}) model~\cite{Loizides:2014vua} 
is employed to generate the volume information in each event (thermal system),
\begin{equation}
	V = h\left[\frac{1-x}{2}n_{\mathrm{part}}+xn_{\mathrm{ncoll}} \right],
\end{equation}
where $n_{\mathrm{part}}$ is the number of participant nucleons and 
$n_{\mathrm{coll}}$ is the number of binary nucleon-nucleon collisions.

With the events generated from the {\tt{MCHRG}} model, the first four cumulants of multiplicity distributions are calculated by~\cite{Adare:2015aqk,Adamczyk:2014fia,Adamczyk:2013dal}
\begin{eqnarray}
	c_{1} &=& \langle N\rangle\equiv M, \nonumber \\
	c_{2} &=& \langle (\Delta N)^{2}\rangle \equiv \sigma^{2}, \nonumber \\
	c_{3} &=& \langle (\Delta N)^{3}\rangle \equiv S\sigma^{3}, \nonumber \\
	c_{4} &=& \langle (\Delta N)^{4}\rangle - 3 c_{2}^{2} \equiv \kappa\sigma^{4}, 
\end{eqnarray}
where $\Delta N = N-\langle N\rangle$ with $N$ being 
the multiplicity of fluctuation measures, and 
$\langle...\rangle$ denotes the event average. Here $M$, $\sigma^{2}$,
$S$ and $\kappa$ are the mean value, variance, skewness and kurtosis 
of the probability distribution. To reduce the centrality bin width 
effect~\cite{Luo:2013bmi}, the cumulants of multiplicity distributions are calculated in each reference multiplicity bin -- the finest
centrality bin in heavy ion experiments. 
The statistical errors are estimated by the Delta theorem~\cite{Luo:2011tp,Luo:2014rea}.

In this work, as an illustrate, I focus on the STAR measurements of net-proton 
distributions and net-charge distributions at $\sqrt{s_{NN}}=39$ GeV. The {\tt MC-Glb} parameters $h= 15.4 fm^{3}$ and $x=0.1$ are determined by the centrality dependence of mean multiplicity of positive and negative charges~\footnote{The mean multiplicity of positive charges ($M_{+}$) and negative charges  ($M_{-}$) are extracted from the data of mean multiplicity of net-charges $M_{Q}$ and the Skellam baselines of $S\sigma$ reported by the STAR collaboration~\cite{Adamczyk:2014fia}, by using the relations $M_{Q}=M_{+}-M_{-}$ and $(S\sigma)_{S}=(M_{+}-M_{-})/(M_{+}+M_{-})$. For more precise constraint, the parameters $x$ and $h$ can be determined by the distribution of reference multiplicity if the data becomes available~\cite{Xu:2016jaz,Xu:2016qzd}.}. The kinetic freeze-out parameters $T_{kin}=120$ MeV and $\beta_{S}=0.6$ are used to include the effect of flow and more realistic simulation of acceptance cuts. The chemical freeze-out parameters $T_{ch}=151$ MeV and  $\mu_{B}=98.5$ MeV are determined by the multiplicity ratios  $p/\pi$ and $p/\bar{p}$, which are consist with the parameters obtained in Ref.~\cite{Alba:2014eba}. Since I do not focus on the details of strangeness fluctuations, the chemical freeze-out parameter $\mu_{S}$ is just obtained from $\mu_{S} = d_{S}/(1+e_{S}\sqrt{s_{NN}})$ with $d_{s}=0.214$ and $e_{S} = 0.161$~\cite{Karsch:2010ck}. The chemical freeze-out parameter $\mu_{Q}$ is determined by the multiplicity ratio of positive charges and negative charges. The mean multiplicity of net-charges is the remainder of two large values, i.e. mean multiplicity of positive and negative charges, which are two orders of magnitude larger than the remainder.
To give a good description of centrality dependence of mean multiplicity of net-charges, therefore, a centrality-dependent $\mu_{Q}$ is parameterized as
\begin{equation}
\mu_{Q} = \mu_{Q0} + a_{Q} \tanh{\left[0.4(\sqrt{n_{\mathrm{part}}}-8)\right]},
\end{equation} 
with $\mu_{Q0}=-4.7$ MeV and $a_{Q}=1.5$ MeV. Besides $\mu_{Q}$, in general, all the chemical and kinetic freeze-out parameters are centrality-dependent for more precise constraint. For simplicity in this work, all of them except $\mu_{Q}$ are set to constants to roughly reproduce the mean multiplicity of positive charges, negative charges, net-charges~\cite{Adamczyk:2014fia}, protons, anti-protons and net-protons~\cite{Adamczyk:2013dal} with some specific acceptance cuts. Note that the main conclusions obtained in the present study are almost independent of the selection of collision energy and  model parameters, and the impact of Glauber parameters on cumulant calculations has been investigated in my previous study~\cite{Xu:2016qzd}.

I apply the same  acceptance cuts as used in experiment~\cite{Adamczyk:2014fia,Adamczyk:2013dal}. More specifically, $|\eta|<0.5$, $0.2<p_{T}<2.0$ GeV for
fluctuation measures ($\pi^{\pm}$, $K^{\pm}$ and $p/\bar{p}$ after removing protons and anti-protons with $p_{T}<0.4$ GeV) and $1.0>|\eta|>0.5$, $0.2<p_{T}<2.0$ GeV for reference particles  (total charged hadrons)
in the net-charge case; $|y|<0.5$, $0.4<p_{T}<0.8$ GeV for fluctuation measures ($p/\bar{p}$) 
and $|\eta|<1.0$, $0.2<p_{T}<2.0$ GeV for reference particles ($\pi^{\pm}$ and $K^{\pm}$ ) in the net-proton case. Here $\eta$ is pseudo-rapidity.

\section{Results and Discussions} 
\label{sec:results}

\begin{figure*}[!hbt]
	\begin{center} 
		\includegraphics[scale=0.28]{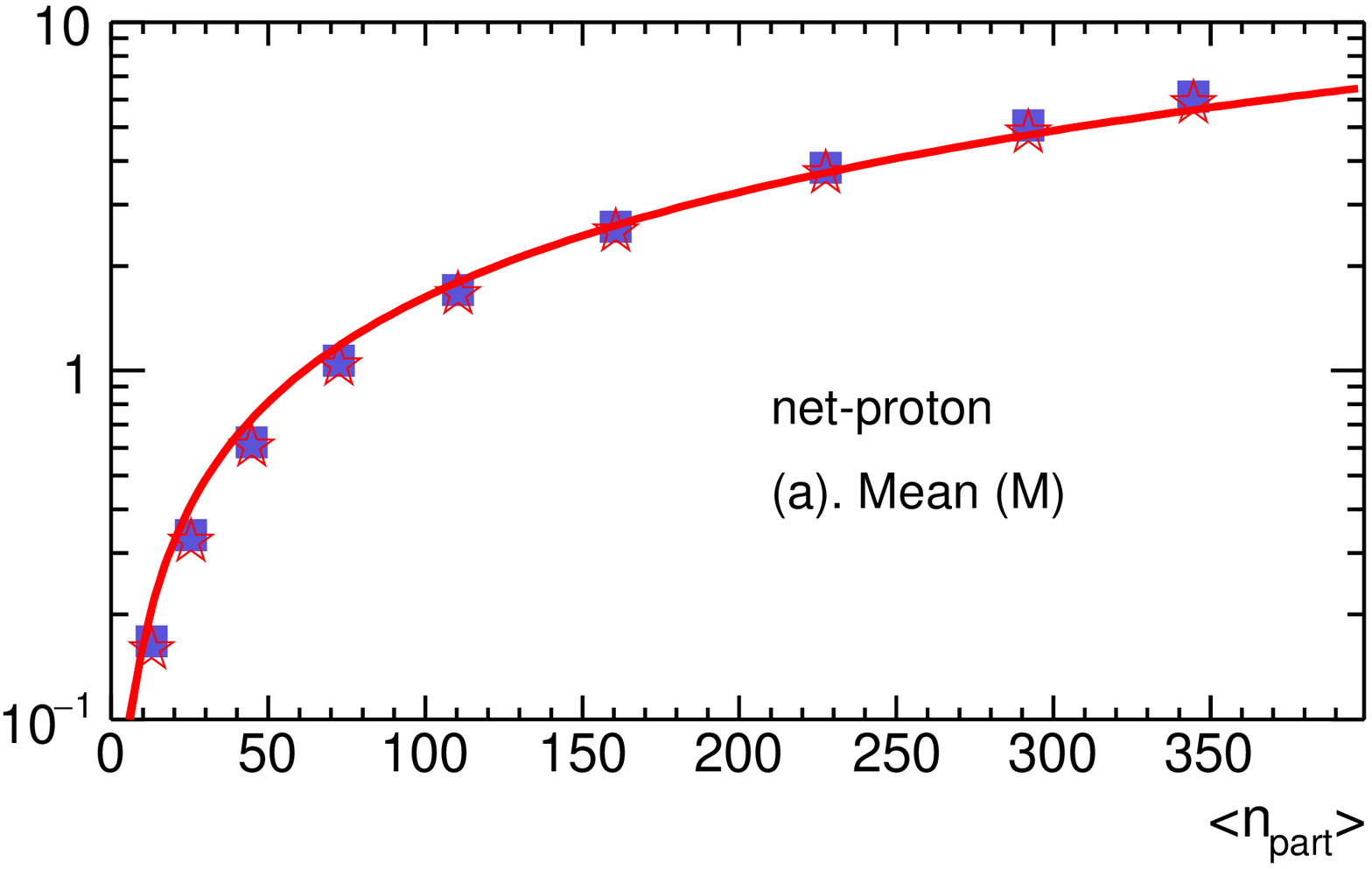}
		\includegraphics[scale=0.28]{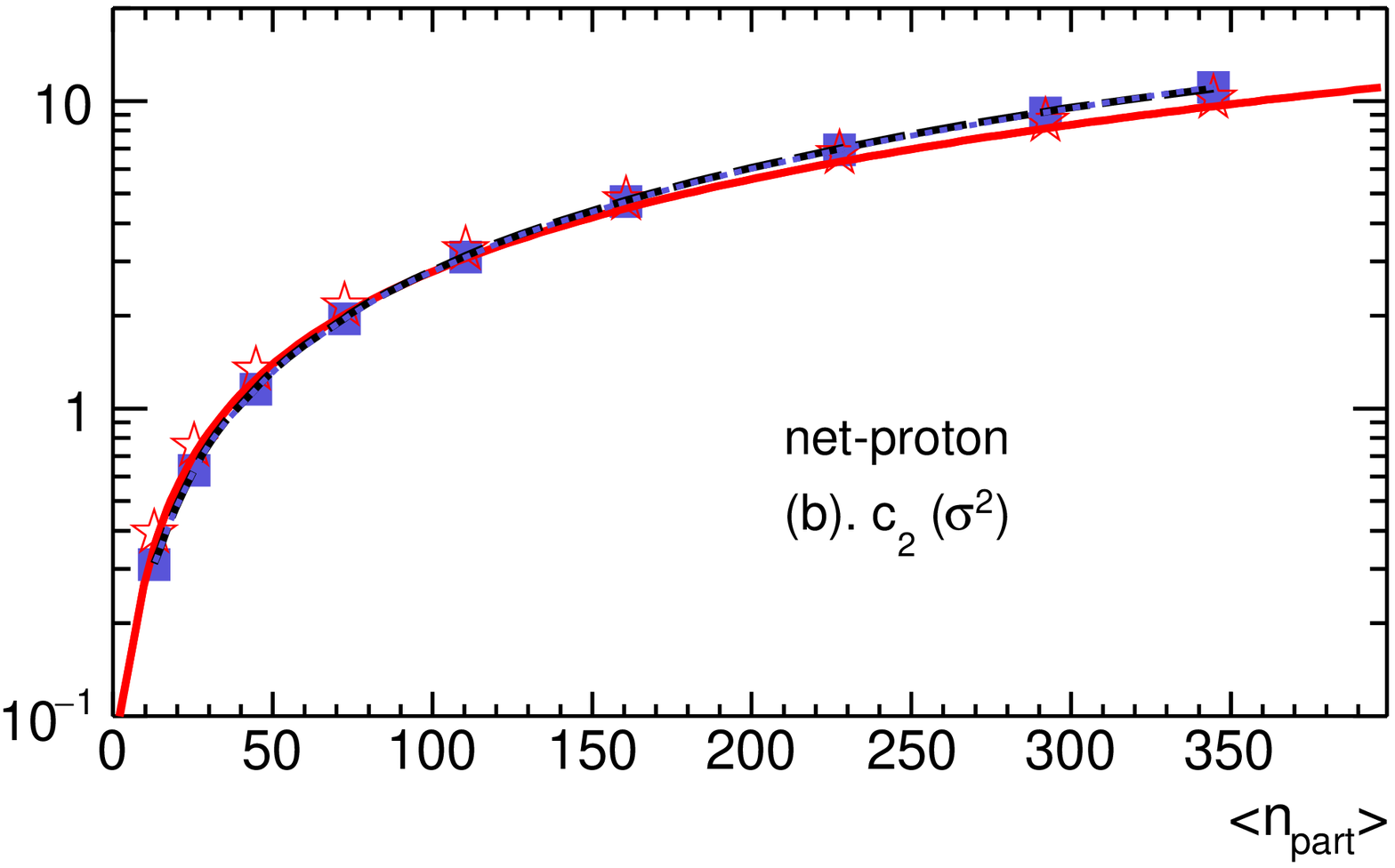} 
		\includegraphics[scale=0.28]{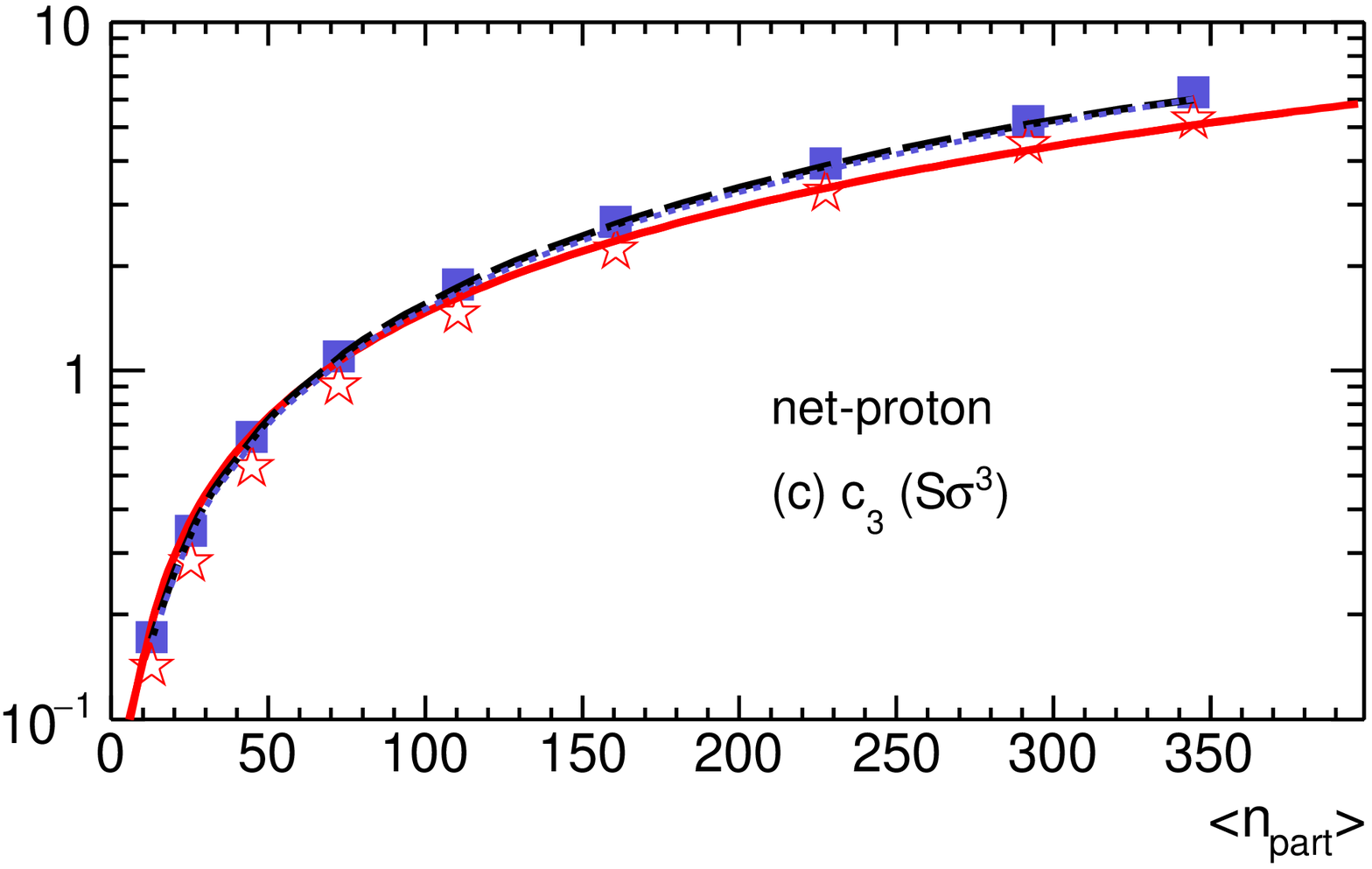} \\
		\includegraphics[scale=0.28]{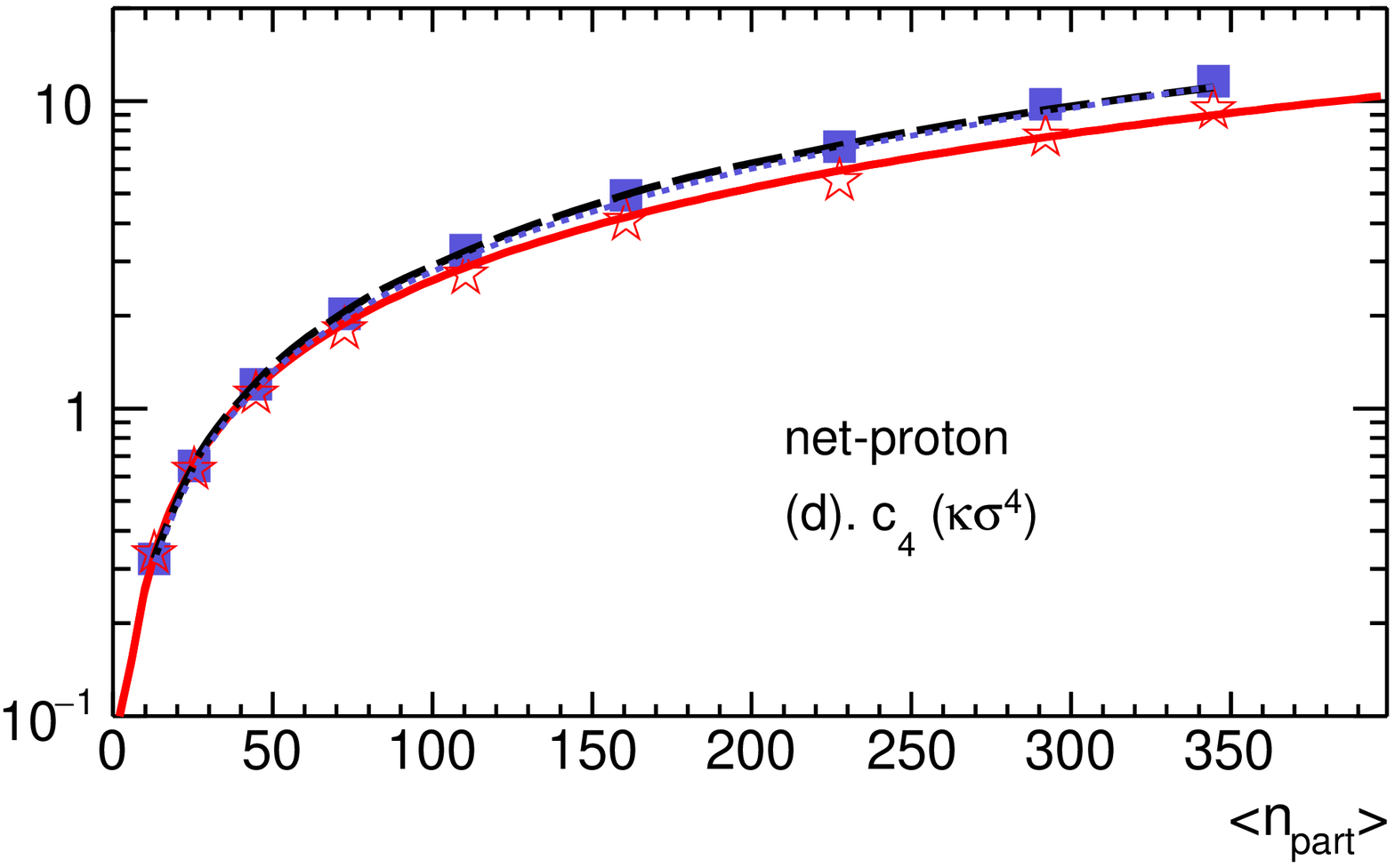}
		\includegraphics[scale=0.28]{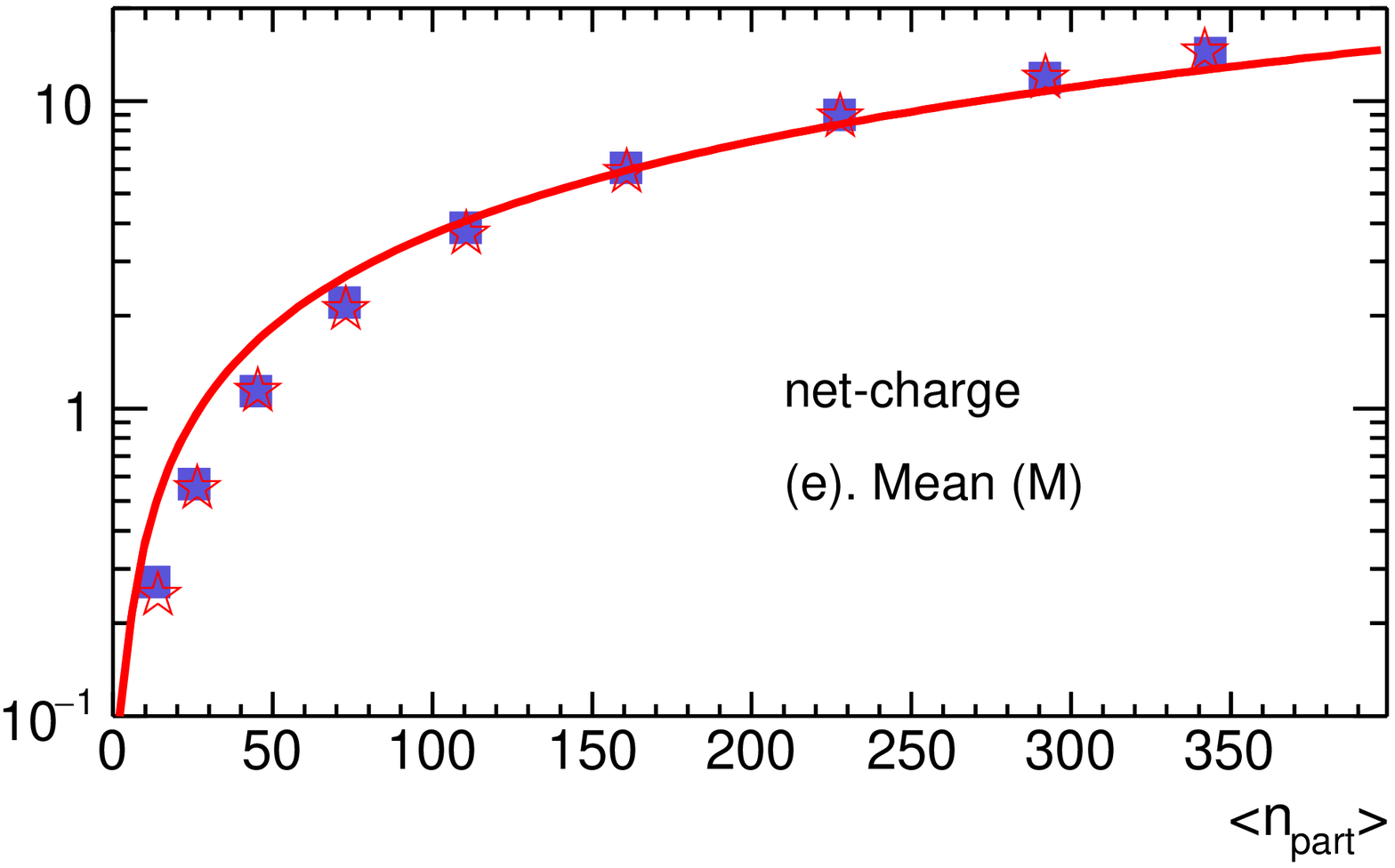}
		\includegraphics[scale=0.28]{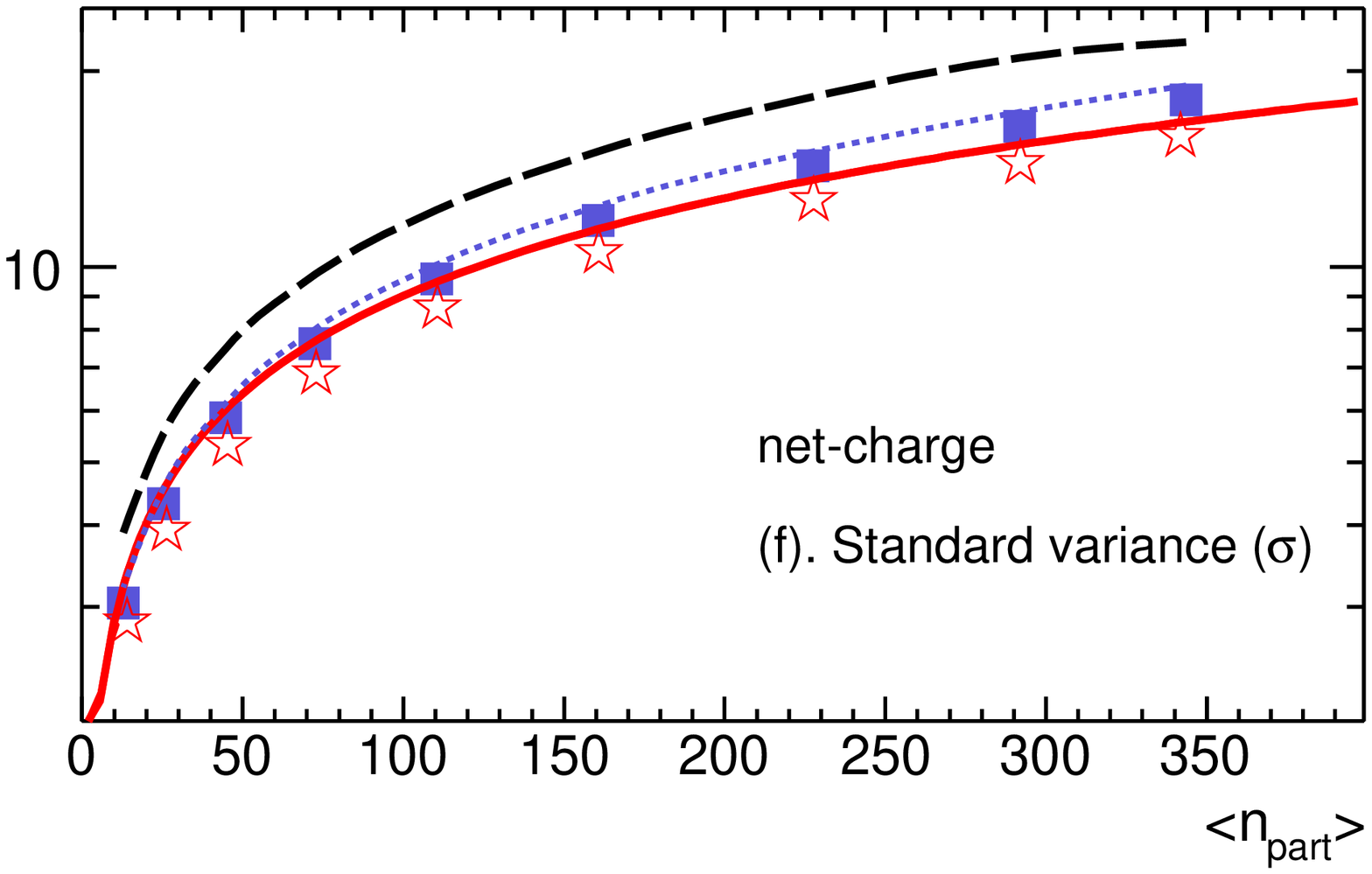}  \\
		\includegraphics[scale=0.28]{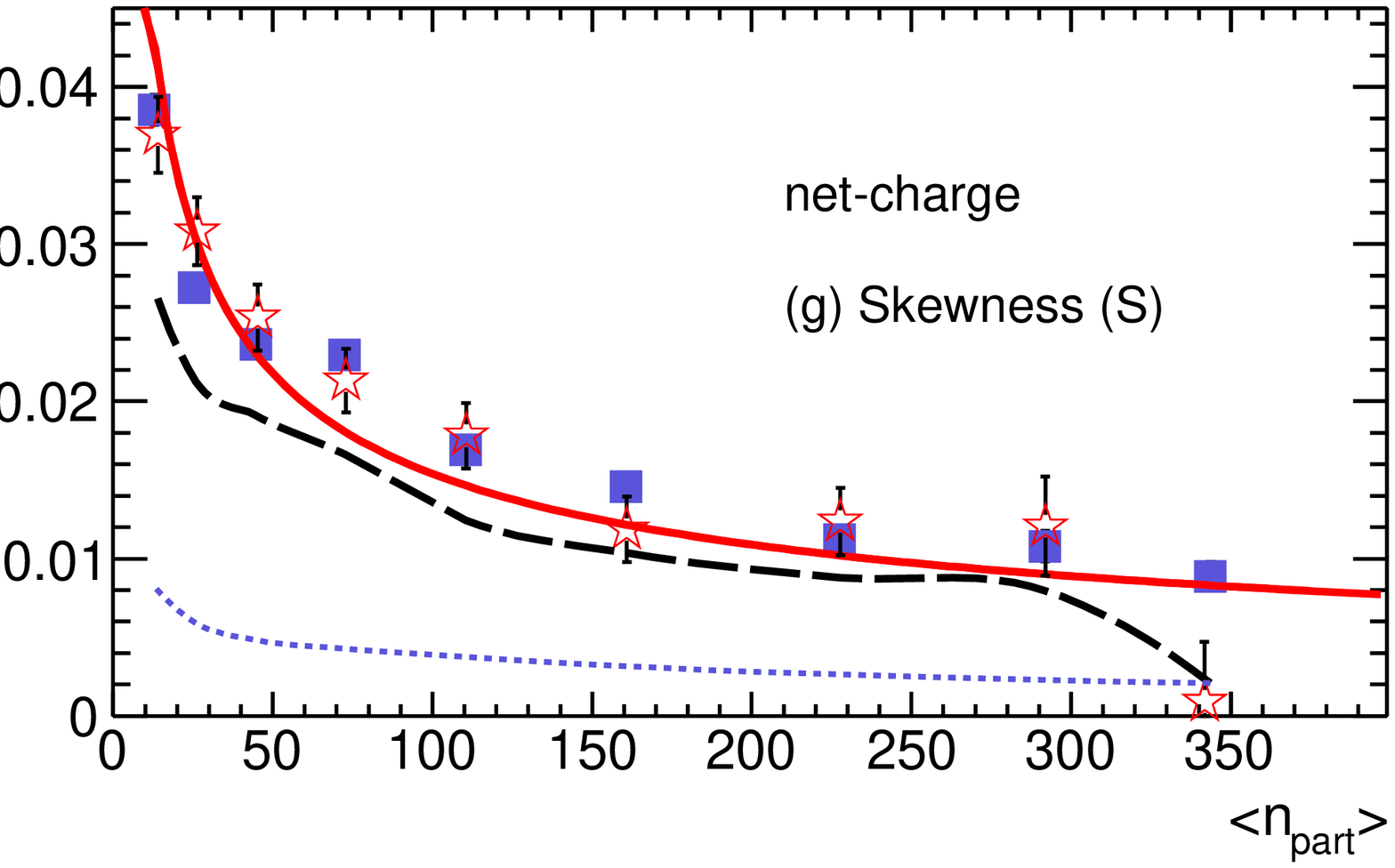}
		\includegraphics[scale=0.28]{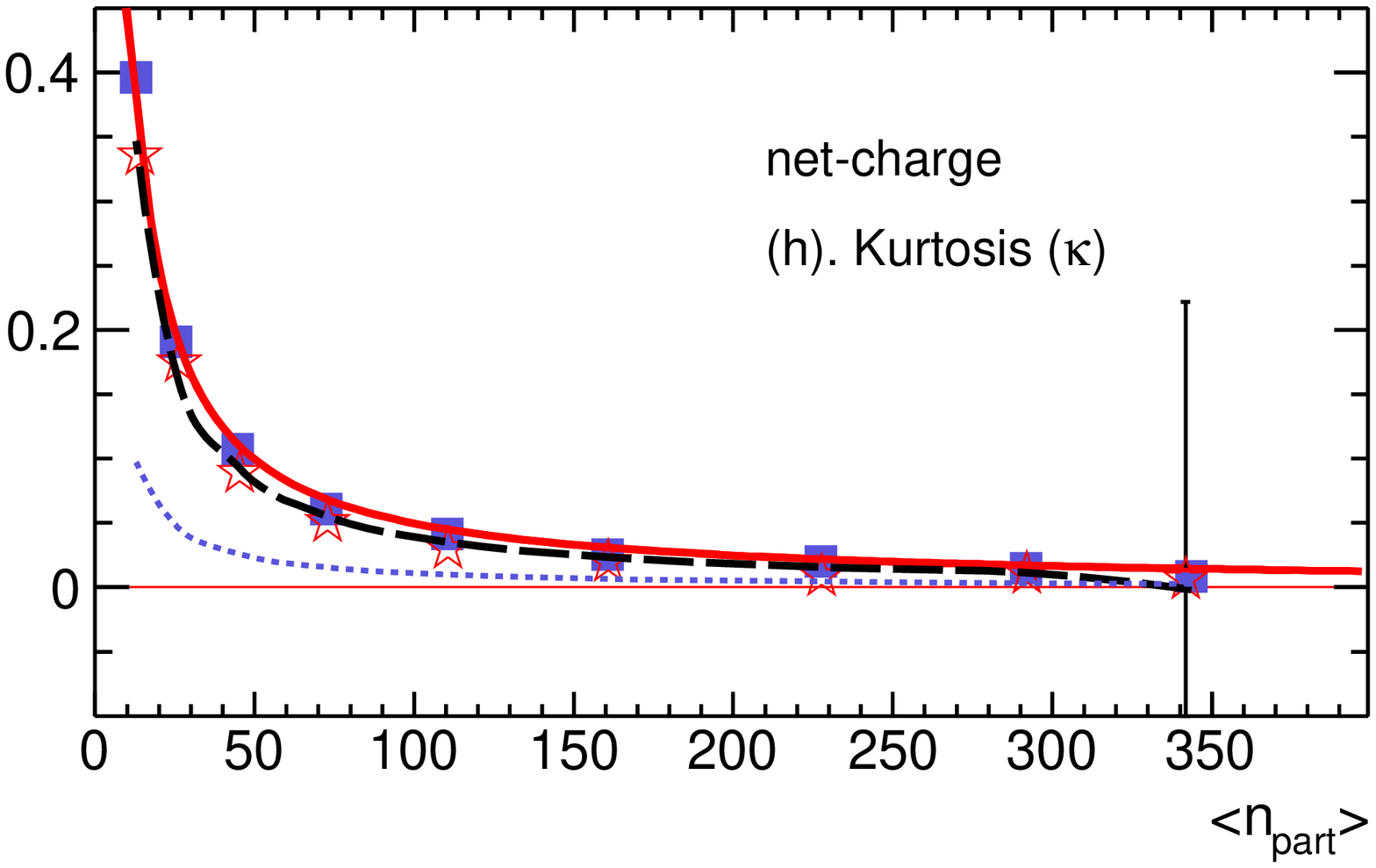}
		\includegraphics[scale=0.28]{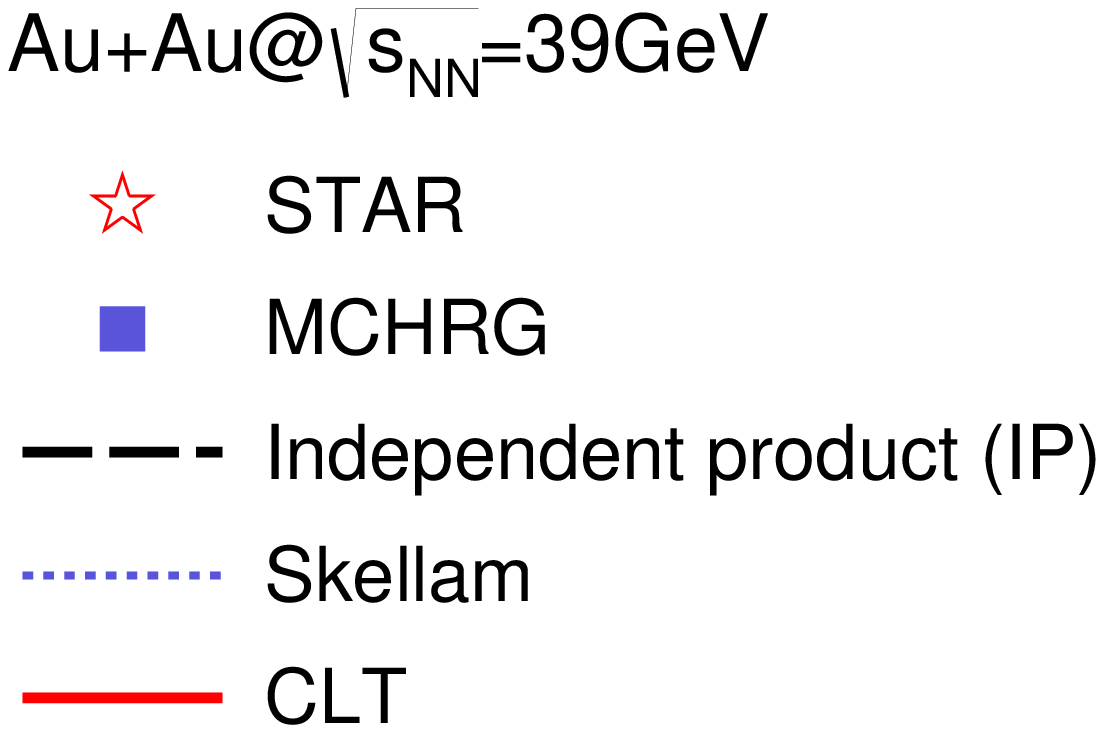}
	\end{center} 
	\caption{(Color online) Centrality dependence of first four cumulants of net-proton distributions and net-charge distributions calculated by the {\tt{MCHRG}} model (blue solid squares).
		The Skellam baselines (blue dotted curves) and independent product (IP, black dashed curves) baselines are obtained from Eq.~(\ref{eq:skellam}) 
		and Eq.~(\ref{eq:IP}), respectively. The data (red open stars) are taken from~\cite{Adamczyk:2014fia} and~\cite{Adamczyk:2013dal}.
	\label{fig:netcharge}} 
\end{figure*}

Fig~\ref{fig:netcharge} shows the centrality-dependent cumulants of net-proton distributions 
and net-charge distributions for Au+Au collisions at $\sqrt{s_{NN}}=39$GeV. To study the effect of volume corrections 
and resonance decays on cumulants of multiplicity distributions, some other baselines are also shown in the figures: one is the 
Skellam baseline, which is obtained by
\begin{equation}
	c_{1}^{\mathrm{S}} = c_{3}^{\mathrm{S}} = c_{1}^{+} - c_{1}^{-},\ \ \ c_{2}^{\mathrm{S}} = c_{4}^{\mathrm{S}} = c_{1}^{+} + c_{1}^{-};
	\label{eq:skellam}
\end{equation}
the other one is the independent product (IP) baseline, which is obtained by
\begin{equation}
	c_{n}^{\mathrm{IP}} = c_{n}^{+} + (-1)^{n}c_{n}^{-}.
	\label{eq:IP}
\end{equation}
Here $c_{n}^{+}$ and $c_{n}^{-}$ are the cumulants of proton (positive charge) distributions and anti-proton
(negative charge) distributions in  the net-proton (net-charge) case calculated by the {\tt{MCHRG}} model. According to the Central Limit Theorem (CLT), the cumulants $c_{n}\propto \langle n_{part}\rangle$, $\sigma\propto\sqrt{\langle n_{part}\rangle}$, $S\propto 1/\sqrt{\langle n_{part}\rangle}$ and 
$\kappa\propto 1/\langle n_{part}\rangle$. The CLT baselines are shown in Fig.~\ref{fig:netcharge} with red-solid curves, which follow the general trend of the {\tt MCHRG} baselines and experimental data.

The {\tt{MCHRG}} baselines, Skellam baselines and IP baselines are almost the same for the first four cumulants of 
net-proton distributions shown in Fig.~\ref{fig:netcharge} (a-d). The reasons are twofold: First, in the absence of critical fluctuations, the volume corrections on 
cumulants of net-proton distributions at $\sqrt{s_{NN}}=39$GeV can be neglected. Second, the resonance decay processes make  
no contributions to the correlation between protons and anti-protons~\cite{Nahrgang:2014fza}. Therefore, for the study of net-proton 
distributions, the {\tt{MCHRG}} model and the HRG model established in Ref.~\cite{Nahrgang:2014fza} are almost the same. 
Note that the {\tt{MCHRG}} model give more  realistic simulations of  acceptance cuts for the fluctuation measures, but it is more computational
expansive due to the high statistics in Monte Carlo simulations. In general the {\tt{MCHRG}} model can reasonably reproduce the data of net-proton distributions,
but, for more precise predictions, more non-critical effects need to be investigated.

The situation is very different for the net-charge fluctuations shown in Fig.~\ref{fig:netcharge}(e-h). The volume corrections play a significant role for the cumulants of net-charge distributions reported by the STAR collaboration. As I have explained in Ref.~\cite{Xu:2016jaz,Xu:2016qzd}, such difference comes from the different magnitude of reduced cumulants~\cite{Xu:2016qzd,Skokov:2012ds} $b\simeq M/(k+1)$ ($M$ is the mean multiplicity of fluctuation measures and $k$ is the corresponding reference multiplicity) in the net-charge case and net-proton case, corresponding to the data reported by the STAR collaboration~\cite{Adamczyk:2014fia,Adamczyk:2013dal}: the reduced cumulants of positive and negative charges are of the order O(1), while the reduced cumulants of protons and anti-protons are much smaller than $1$. The {\tt MCHRG} baselines support my conclusions about volume corrections on multiplicity distributions given in Ref.~\cite{Xu:2016jaz,Xu:2016qzd} with a simple statistical model. Meanwhile, the resonance decays and resulting correlations between positive charges and negative charges are included in the {\tt MCHRG} model. Therefore, Monte Carlo simulations are very appropriate for the study of net-charge fluctuations than a simple statistical model.

The large gaps between the Skellam baselines and IP baselines indicate that the effects of volume corrections and resonance decays on 
cumulants of positive charge distributions and negative charge distributions are important. Analogously, the deviations of the {\tt MCHRG} 
baselines from Skellam baselines indicate that the effect of volume corrections and resonance decays on skewness and kurtosis 
of net charge distributions are important, see Fig.~\ref{fig:netcharge}(g,h). However, I will show that the effect of volume corrections on the 
variances of net-charge distributions can be neglected, which make the {\tt MCHRG} baselines close to the Skellam baselines, 
see Fig~\ref{fig:netcharge}(f). The deviations of the {\tt MCHRG} baselines from the IP baselines indicate strong correlations between 
positive charges and negative charges. With the volume corrections, resonance decays, and more realistic simulations of acceptance cuts,  the {\tt{MCHRG}} 
model can reasonably reproduce the skewness and kurtosis of net-charge distributions reported by the STAR collaboration~\cite{Adamczyk:2014fia}, 
but shown obvious deviations for the variances. The results indicate that the correlations between positive and negative charges from other sources are required to quantitatively reproduce the data of variances of net-charge distributions, which are beyond the scope of the {\tt MCHRG} model proposed in this work. The skewness and kurtosis of net-charge distributions seem insensitive to these correlations and its deviations from Skellam baselines are dominated by the effect of volume corrections.

\begin{figure}[!hbt]
	\begin{center} 
		\includegraphics[scale=0.4]{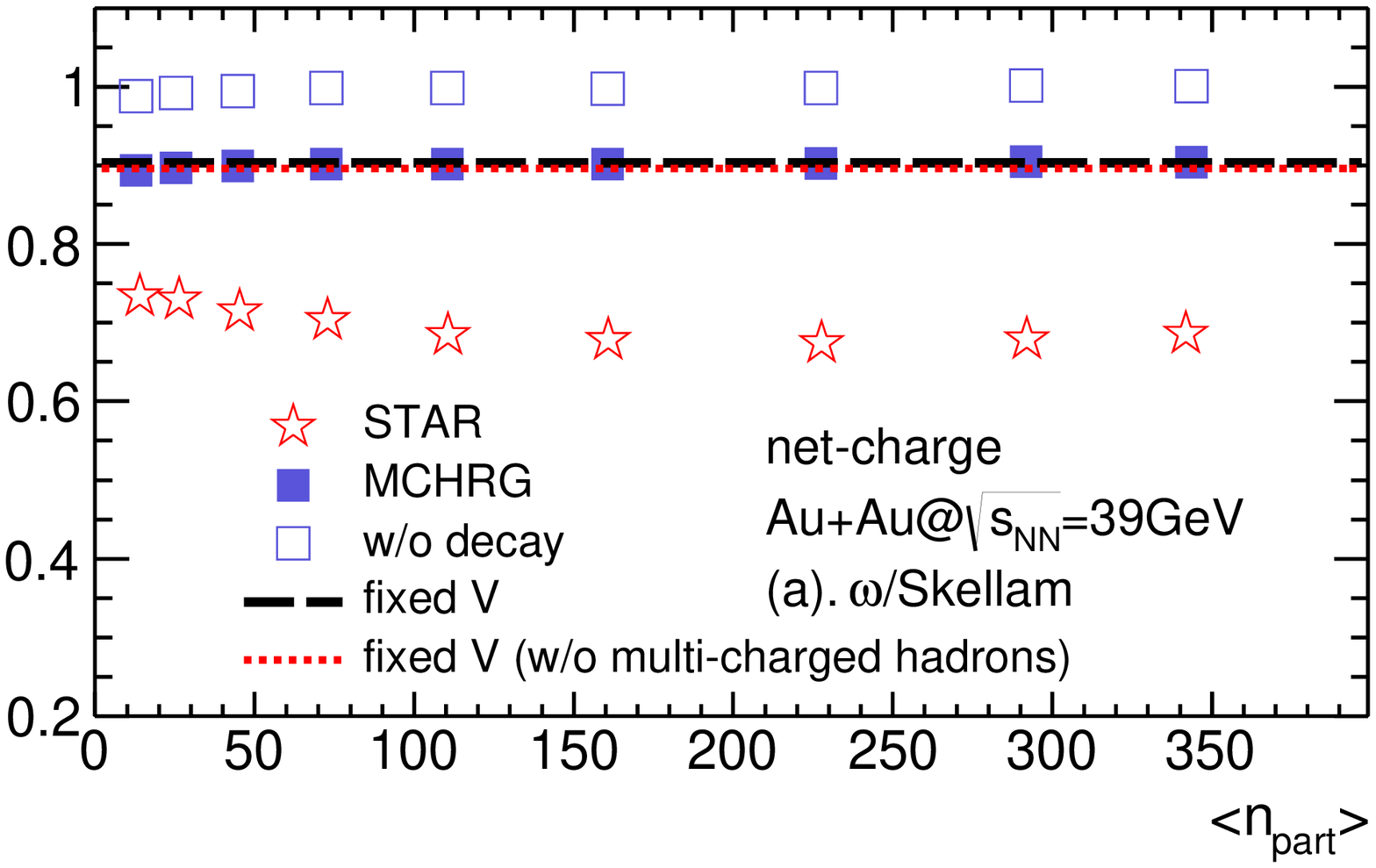} \\
		\includegraphics[scale=0.4]{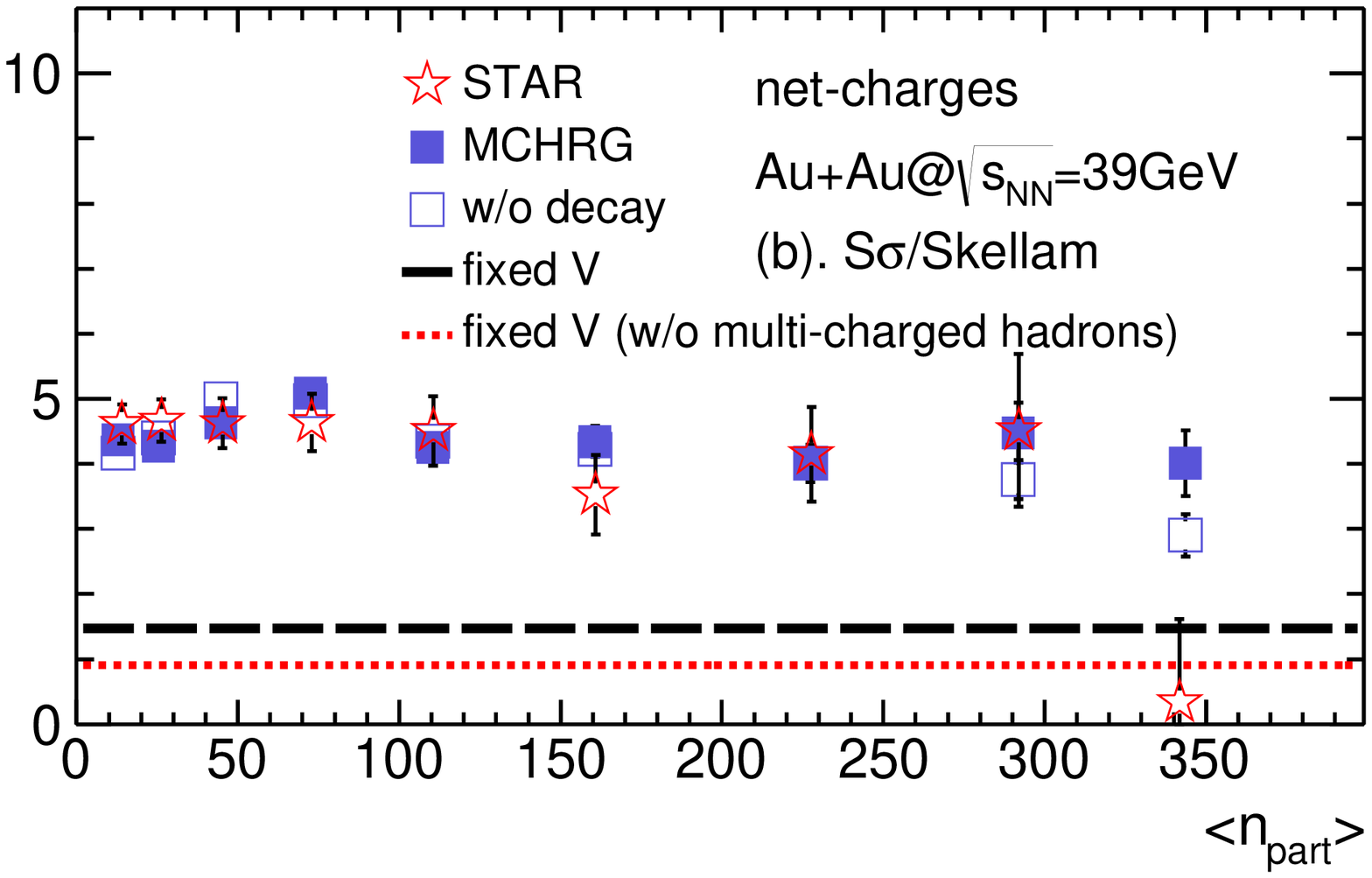} \\
		\includegraphics[scale=0.4]{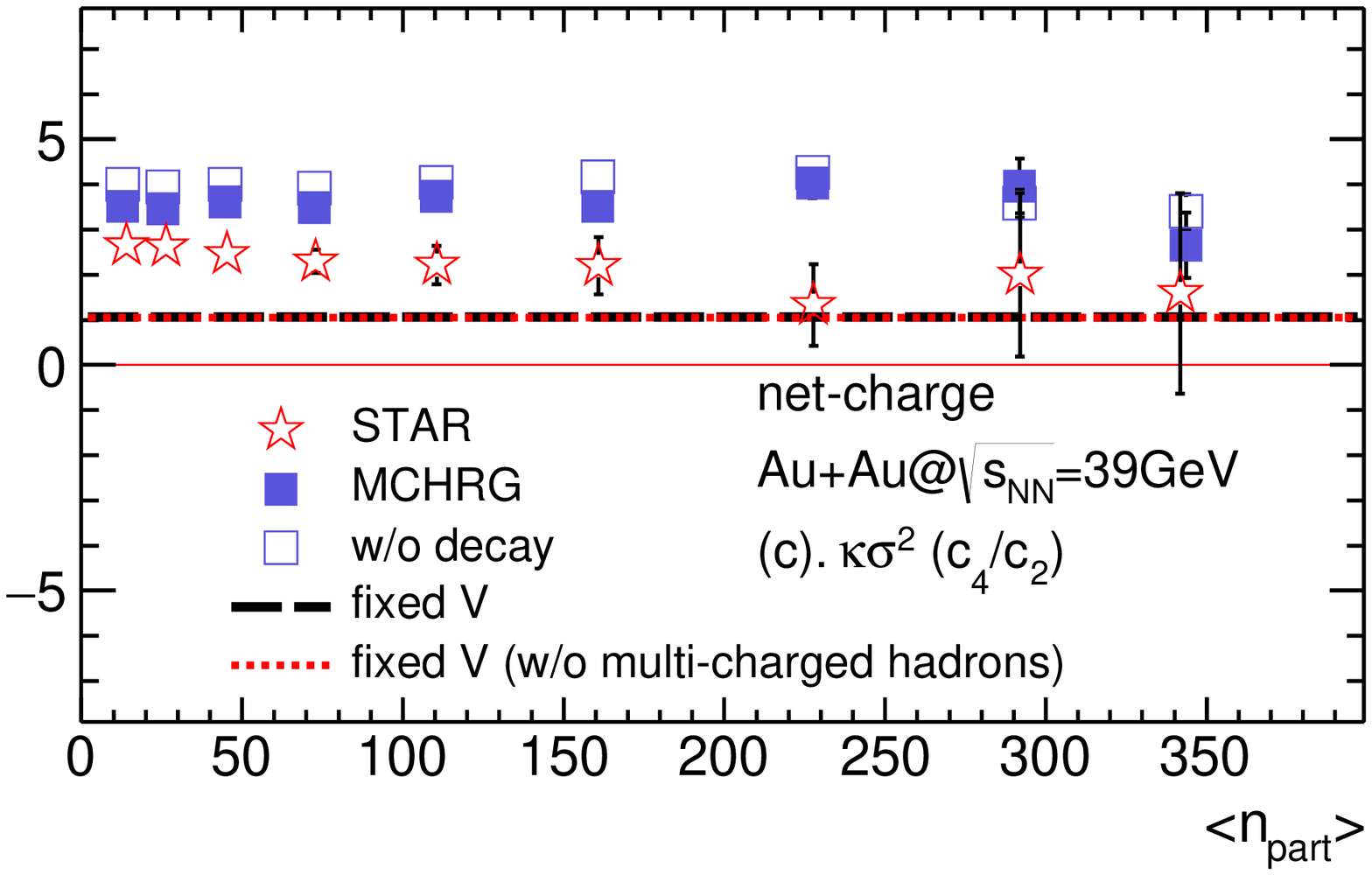}
	\end{center} 
	\caption{(Color online) Centrality-dependence of cumulant products $\omega$, $S\sigma$ and $\kappa\sigma^{2}$ of net-charge distributions from the full {\tt MCHRG}+{\tt MC-Glb} simulations(blue solid squares),{\tt MCHRG}+{\tt MC-Glb} simulations without resonance decays(blue open squares) and {\tt MCHRG} simulations in a fixed volume with (black dashed lines) and without (red-dotted lines) multi-charged hadrons. The data (red open stars) are taken from~\cite{Adamczyk:2014fia}.
	\label{fig:cumulantp}} 
\end{figure}

To explore in depth the effects of volume corrections and resonance decays on the cumulants of net-charge distributions individually, I further calculate the cumulant products 
\begin{equation}
	\omega=c_{2}/c_{1},\ \ S\sigma=c_{3}/c_{2},\ \ \kappa\sigma^{2}=c_{4}/c_{2}
\end{equation}
in  three different cases:
(1) {\tt MCHRG} simulations with {\tt MC-Glb} volume distributions discussed before. (2)  Similar to case (1) but for the  primordial particles
before resonance decays, (3) {\tt MCHRG} simulations in a fixed 
volume with volume $V=1540 fm^{3}$  and $\mu_{Q}=\mu_{Q0}=-4.7$ MeV. Obviously, case (2) only include the effect of volume corrections, while
case (3) only include the effect of resonance decays and the latter cumulants are independent of centrality.

The results are shown in Fig.~\ref{fig:cumulantp}. The volume corrections on $\omega$ of net-charge distributions can be neglected~\cite{Xu:2016csh}, though the volume corrections on $\omega$ of positive charges and negative charges are important~\cite{Xu:2016jaz,Xu:2016qzd}. This is because the mean multiplicity of net-charge distribution is much smaller than the reference multiplicity, though the magnitudes of mean multiplicity of positive and negative charges are the same order of reference multiplicity.  The volume corrections play 
significant role for $S\sigma$ and $\kappa\sigma^{2}$, which make their values deviate far away from the Skellam predictions. The resonance decays make $\omega$ of net-charge distributions smaller than the Skellam predictions, but it make $S\sigma$ and $\kappa\sigma^{2}$ larger than the Skellam predictions. From the {\tt MCHRG} calculations, I find that the deviations of $\omega$ from Skellam distributions are mainly due to the effect of resonance decays, while the deviations of $S\sigma$ and $\kappa\sigma^{2}$ are mainly due to effect of volume corrections.

For the effect of resonance decays, the multi-charged hadrons play special role in study of net-charge distribution~\cite{Ejiri:2005wq}. To identify the effect of multi-charged hadrons on net-charge distributions through the resonance decay process, I then calculate the cumulant products of net-charge distributions in case (3) without the decay channels of multi-charged hadrons. The results are shown in red-dotted curves of Fig.~\ref{fig:cumulantp}. The resonance decays of multi-charged hadron enhance the fluctuations of net-charges, as it have been investigated in Ref.~\cite{Ejiri:2005wq}. Comparison the results with (black-dashed lines) and without (red-dotted lines) multi-charged hadrons in Fig.~\ref{fig:cumulantp}, I find that the effect of multi-charged hadrons can be neglected for $\omega$ and $\kappa\sigma^{2}$ of net-charge distributions, while they make substantial contributions to $S\sigma$ of net-charge distributions.

The deviations of $\omega$, $S\sigma$ and $\kappa\sigma^{2}$ from data 
are mainly due to the fact that the {\tt MCHRG} model fail to quantitatively
reproduce the variances of net-charge distributions(see Fig.~\ref{fig:netcharge}(f)).
The results imply that, for more realistic baselines predictions of the first
four cumulants of net-charge distributions, some other effects are expected to
make relevant contributions to the variances, without significant affecting
its skewness and kurtosis.

\section{Conclusions} 
I investigated the cumulants of  net charge distributions and  net-proton
distributions within a Monte Carlo hadron resonance gas ({\tt{MCHRG}}) model. 
To study the centrality dependence of multiplicity distributions, as well as the 
effect of volume corrections on its cumulants, the volume distributions are generated by 
a Monte Carlo Glauber ({\tt MC-Glb}) model. For the net-proton distributions, even with more realistic
simulations of acceptance cuts and volume corrections, the {\tt MCHRG} calculations are consist 
with the semi-analytical calculations given in Ref.~\cite{Nahrgang:2014fza}. 
However, both the effect of volume corrections, 
resonance decays, as well as the resulting correlations between positive 
charges and negative charges, that are important for the cumulants of net-charge distributions. With these effect, 
the {\tt MCHRG} calculations provide more realistic baseline predictions for the  cumulants of net-charge distributions than previous 
HRG studies.

Except the variances of net-charge distributions, the {\tt MCHRG} model can
reasonably explain the cumulants of net-proton distributions and
net-charge distributions reported by the STAR collaboration. To explore in
depth the effect of volume corrections and resonance decays on the cumulants
of net-charge distributions individually, I also calculated the cumulant
products of net-charge distributions with primordial particles, as well as the
cumulant products of net-charge distributions in a fixed volume with and without multi-charged hadrons. 
The deviations of $\omega$ of net-charge distributions from Skellam
expectations are mainly due to the effect of resonance decays, while the 
deviations of $S\sigma$ and $\kappa\sigma^{2}$ are mainly due to  the effect of volume corrections.

Note that, for more realistic baseline predictions in the future, more additional effects beyond 
the non-interacting {\tt MCHRG} model used in this work are in order, e.g. the correlations between 
positive charges (protons) and negative charges (anti-protons) from  the charge (baryon)
conservation laws~\cite{Bzdak:2012an}, the dynamic evolution in hadronic phase~\cite{Steinheimer:2016cir},
etc. The results in the present study imply that these effects are expected to
make substantial contributions to the variances of net-charge distributions, 
without significant affecting its skewness and kurtosis.

\section*{Acknowledgments}  
This work is supported by the China Postdoctoral Science Foundation under
Grant No.~2015M580908.

\bibliographystyle{elsarticle-num}
\bibliography{ref}

\end{document}